\documentclass[11pt]{article}   
\usepackage{amsthm, amsmath}   
\usepackage{graphicx}
\theoremstyle{plain}

\theoremstyle{definition}

\theoremstyle{remark}

\title{\bf The Accelerated Euclidean Algorithm}
\author{Sidi Mohamed Sedjelmaci}
\date{}
\begin{document}
\maketitle

\begin{abstract}
We propose  a new GCD algorithm called Accelerated Euclidean 
Algorithm, or AEA for short, which matches 
the  $O(n \log^2 n \log \log n)$ time complexity of Sch\"onhage algorithm for 
$n$-bit inputs.
This new GCD algorithm is designed for both integers and 
polynomials. We only focus our study to the integer case, the polynomial case 
is currently addressed~($[3]$). 
\end{abstract}
\section{Introduction} \label{int}
The algorithm is based on a half-gcd like procedure, but unlike Sch\"onhage's 
algorithm, it is iterative and therefore avoids the 
penalizing calls of the repetitive recursive procedures. The new half-gcd 
procedure reduces the size the integers 
at least a half word-memory bits per iteration only in single precision. 
By a dynamic updating process, we obtain 
the same recurrence and the same time performance as in the Sch\"onhage approach. 

Throughout, the following notation is used. $W$ is a word memory, i.e.: $W=16$, 
$32$ or $64$.  
Let $u\ge v>2$ be positive integers where $u$ has a $n$ bits with $n \geq 32$.  
Given a non-negative integer 
$x \in N$, $\ell(x)$ represents 
the number of its significant 
bits, not counting leading zeros, i.e.:~$\ell(x)= \lceil \log_{2}(x+1)\rceil$.  
 For the sake of readability integers $U$ and $V$ will be 
represented as 
a concatenation of $l$ sets of $W$ bits integers (except for the last 
set which may be shorter), with $l= \lceil  n/W \rceil $, i.e.: if  
$U=  \sum_{i=0}^{l-1}2^{iW}~U_{l-i} \quad {\rm with} 
\quad U_1 \neq 0$ and $V=  \sum_{i=0}^{l-1}2^{iW}~V_{l-i}$, 
then we write symbollically 
$U=U_1 \bullet U_2 \ldots \bullet U_l$ and 
$V=V_1 \bullet V_2 \ldots \bullet V_l$.\\
\noindent We  use 
specific vectors which represent interval subsets from of $U$ and $V$, 
by:\\
\hspace*{1.5cm} $X[i..j]=       
\left( \begin{array}{c} U_i \bullet U_{i+1} \bullet \dots \bullet  U_j \\ 
	V_i \bullet V_{i+1} \bullet \dots \bullet V_j \end{array} \right)
	 \ ; \ X[i]=       
\left( \begin{array}{c} U_i  \\ 
V_i \end{array} \right) \ ; \ {\rm for} \ 1 \leq i<j \leq l$.\\	 
Let  $M(x)$ be the cost of a multiplication of two $x$-bit integers. 
The function $M(x)$ depends on the algorithm used to carry out 
the multiplications. The fast Sch\"onhage-Strassen  
algorithm~($[6]$) performs these multiplications in 
$M(x)= O(n \log n \log \log n)$. 

\section{AEA: The Accelerated Euclidean Algorithm}  
 The following algorithm is  based on two main ideas:\\
\noindent \hspace*{.5cm} $\bullet$ The computations are done in a Most 
Signifivant digit First (MSF) way.\\
\hspace*{.5cm} $\bullet$ Update by multiplying, with the current matrix, 
 {\tt ONLY} twice the number 
of leading 
\hspace*{.8cm} bits we have already chopped, 
{\tt NOT} the others leading bits.\\
\newpage
\noindent Algorithm {\tt AEA}. 
\begin{quote}
{\bf Input~:} $u \ge v > 2$ with $ u \geq 8W$;  ~ $n= \ell(u)$; \\
{\bf Output~:} a $2 \times 2$ matrix $M$ and $(U',V')$ 
such that \\
$M \times (U,V)= (U',V')$ and $ \ell(V') 
\leq  \ell(U) - 2^{ \lfloor \log_2(n/W) \rfloor -1} W$. \\
\noindent {\bf Begin}\\
{\scriptsize 1.}  \hspace*{.4cm} $(U,V) \leftarrow (u,v)$; 
$s\leftarrow  \lfloor \log_2(n/W)  \rfloor $;\\
{\scriptsize 2.} \hspace*{.4cm} {\bf if} 
$\ell(V) \leq  \lceil \ell(U)/2 \rceil +1$ { \bf return} $I$;\\
{\scriptsize 3.} \hspace*{.4cm} {\bf if} 
$\ell(V) > \lceil \ell(U)/2 \rceil +1$  \\
{\scriptsize 4.} \hspace*{.4cm} {\bf For} $i=1$  {\bf to} $2^{s-1}$  \\ 
{\scriptsize 5.} \hspace*{1.2cm} {\bf if $U_i = 0~or~V_i \neq 0$} (Regular case)\\
{\scriptsize 6.} \hspace*{1.7cm}  $h \leftarrow 0$; \\
{\scriptsize 7.} \hspace*{1.7cm}  {\bf if} ($i$ odd)  
$L_0 \leftarrow {\tt ILE}(X[i..i+1])$; {\bf update$_L$}$(i,h)$; \\
{\scriptsize 8.} \hspace*{1.7cm}  {\bf else} \\
{\scriptsize 9.} \hspace*{2.1cm} $R_0 \leftarrow {\tt ILE}(X[i..i+1])$; 
{\bf update$_R$}$(i,h)$; \\
{\scriptsize 10.} \hspace*{2cm} $x \leftarrow i/2; \ h \leftarrow h+1$; \\ 
{\scriptsize 11.} \hspace*{2cm} {\bf While} ($x$ even)  \\ 
{\scriptsize 12.} \hspace*{3cm} $R_h \leftarrow  R_{h-1} \times L_{h-1}$; 
{\bf update$_R$}$(i,h)$; \\ 
{\scriptsize 13.} \hspace*{3.1cm}  $x \leftarrow x/2; \ h \leftarrow h+1$; \\
{\scriptsize 14.} \hspace*{2.1cm} {\bf Enwhile} \\
{\scriptsize 15.} \hspace*{2.1cm}  $L_h \leftarrow  R_{h-1} \times L_{h-1}$; 
{\bf update$_L$}$(i,h)$; \\
{\scriptsize 16.} \hspace*{1.6cm}  {\bf Endelse} \\
{\scriptsize 17.} \hspace*{1cm} {\bf else} {\tt Irregular(i)} 
($U_i \neq 0~and~V_i = 0$) ;  \\
{\scriptsize 18.} \hspace*{.4cm} {\bf EndFor} \\
{\scriptsize 19.} \hspace*{.4cm} {\bf Return} $L_h$ and $(U,V)$;\\
{\bf End} 
\end{quote}  
The algorithm {\tt ILE} (borrowed from~$[7]$) runs the extended Euclidean algorithm and stops when the
remainder has roughly the half size of the inputs.\\	
Algorithm {\tt ILE}. 
\begin{quote}
{\bf Input~:} $u_0 \ge u_1 \ge 0$   \\
{\bf Output~:} a $2 \times 2$ matrix $M$ and $(u_{i-1},u_i)$ such that 
$M \times (u_0,u_1)= (u_{i-1},u_i)$ and $ \ell(u_i) \leq \frac{1}{2} \ell(u_0)$. \\
{\bf Begin}\\
{\scriptsize 1.} $n= \ell(u_0)$; \ $p= \ell(u_1)$;\\
{\scriptsize 2.} {\bf if} $p < \lceil n/2 \rceil +1$ 
        {\bf return} $M= 
	\left( \begin{array}{cc} 1 & 0 \\ 
	0 & 1 \end{array} \right) $;\\
{\scriptsize 3.} {\bf if} $p \geq \lceil n/2 \rceil +1  $ \\	
{\scriptsize 4.} \quad $m=p- \lceil n/2 \rceil -1     $;\\
{\scriptsize 5.} \quad Apply Extended Euclid Algorithm until \  
          $|a_i| \leq 2^m < |a_{i+1}|$;\\
{\scriptsize 6.} \quad  {\bf return} 
	$M= 
	\left( \begin{array}{cc} a_{i-1} & b_{i-1} \\ 
	a_i & b_i \end{array} \right) $ \ and  \ $(u_{i-1}, u_i)$.\\
{\bf End}.
\end{quote}
The functions {\bf update$_L$} or {\bf update$_R$} update not all the bits of 
the operands, 
but only the next useful small vectors, in order to get the next needed matrix. 
The irregular case is when  
$U_i \neq 0$ and $V_i = 0$, i.e.:  
one or many components are all equal to zero. Roughly speaking, we 
perform an euclidean division in order to make an efficient reduction 
and continue our process. The aim is to preserve, at most, the general
scheme of our process. The basic idea is to use full updated vectors, i.e.: 
vectors updated with all the previous matrices $L_h$. 


\section{An Example} \label{exa}
In order to carry out single-precision computations we take $W=20$. Recall 
that the notation  
$ \left( \begin{array}{c} A \\ 
	B \end{array} \right) =$
$ \left( \begin{array}{c} a ~\bullet ~c \\ 	
	b ~\bullet ~d \end{array} \right)$ means 
$ \left( \begin{array}{c} A \\ 
	B \end{array} \right) =$	
$ \left( \begin{array}{c} a \\ 
	b \end{array} \right) \times 2^W + $	
$ \left( \begin{array}{c} c \\ 
	d \end{array} \right) $, where $A$, $B$, $a$, $b$, $c$ and $d$ 
	are integers (c.f. notation in Section~\ref{int}).\\

\noindent Let $
\left( \begin{array}{c} u \\ 
	v \end{array} \right) =
\left( \begin{array}{c} 922375420941 \\
	7075 9930  7587 \end{array} \right)$, then, with our notation, 
	we obtain \\
$ \left( \begin{array}{c} U \\ 
	V \end{array} \right) =	 
 \left( \begin{array}{c} u \\ 
	v \end{array} \right) =	
\left( \begin{array}{c} 879645 \\ 
	674819 \end{array} \right) \times 2^{20}+
\left( \begin{array}{c} 785 421\\ 
	299 843 \end{array} \right)= 		
\left( \begin{array}{c} 879645 ~\bullet ~785 421 \\ 
	674819 ~\bullet ~299 843 \end{array} \right)$.\\

We have $U_1=879645$ and $V_1=674819$ then $\ell(U_1)= \ell(V_1)=20$, 
$n=p=20$ and $m_1=p-1 - \lceil \frac{n}{2}  \rceil =9 $. 
Thus, in order to compute $ILE(U_1,V_1)$, we must stop the Extended 
Euclid Algorithm at $|a_i| \leq 2^9$. We obtain the matrix ( $|a_i|<2^9=512$)\\
\hspace*{3.5cm} $N_1=ILE(U_1,V_1)=   
	\left( \begin{array}{cc} 369 & -481 \\ 
	-425 & 554 \end{array} \right) \quad  {\rm then}: $\\	
$N_1 \left(\begin{array}{c} U\\ 
	V \end{array}\right) = 
N_1 \left(\begin{array}{c} U_12^{20}+U_2\\ 
	V_12^{20}+V_2 \end{array}\right) = 
 N_1 \left(\begin{array}{c} U_1\\ 
	V_1 \end{array}\right) 2^{20}+
N_1 \left(\begin{array}{c} U_2\\ 
	V_2 \end{array}\right)$. \\
	\noindent 
So $N_1 \times 
\left( \begin{array}{c} U_1 2^{20}+U_2 \\ 
	V_1 2^{20}+V_2 \end{array} \right) =	
\left( \begin{array}{c} 1066 \\ 
	601 \end{array} \right) \times 2^{20} +
\left( \begin{array}{c} 138 \\ 
	-160 \end{array} \right) \times 2^{20} + 
\left( \begin{array}{c} 892378 \\ 
	81257 \end{array} \right) $\\		
\hspace*{3.5cm} $ = \left( \begin{array}{c} 1204 \\ 
	441 \end{array} \right) \times 2^{20} +
\left( \begin{array}{c} 892378 \\ 
	81257 \end{array} \right)$,\\	
\noindent hence\\
$ \left( \begin{array}{c} U_1 \\ 
	V_1 \end{array} \right)	= 
\left( \begin{array}{c} 1204 \\ 
	441 \end{array} \right)$ and 	
$ \left( \begin{array}{c} U_2 \\ 
	V_2 \end{array} \right)	= 
\left( \begin{array}{c} 892378 \\ 
	81257 \end{array} \right)$.\\
		
Now we can apply {\tt ILE} to the new integers (less than $32$ bits) 
$U=1263377882$ and $V=462503273$. We have $\ell(U)= 31$ and $\ell(V)
=29$. We obtain $m=8$ and the second matrix (in $32$-bits single 
precision) \\
\hspace*{3.5cm} $N_2=ILE(U,V)=   
	\left( \begin{array}{cc} -41 & 112 \\ 
	231 & -631 \end{array} \right) $ \\
and $M= N_2 \times N_1= \left( \begin{array}{cc} -41 & 112 \\ 
	231 & -631 \end{array} \right) \times 
\left( \begin{array}{cc} 369 & -481 \\ 
	-425 & 554 \end{array} \right)= 	
\left( \begin{array}{cc} -62729 & 81769 \\ 
	353414 & -460685 \end{array} \right)$.\\

\noindent Moreover, at this level, we may consider that $U_1$ and $V_1$ 
are 
eliminated (chopped), even if $U_2$ has some extra bits,  and that $U_2$ 
and $V_2$ 
are updated as follows:\\

$ \left( \begin{array}{c} U_2 \\ 
	V_2 \end{array} \right)	= 
\left( \begin{array}{c} 1873414  \\ 
	725479 \end{array} \right) = 
\left( \begin{array}{c} 1 \bullet 824838  \\ 
	0 \bullet 725479 \end{array} \right) $;
	 \ \ $\ell(U_2)=21$; \ \ $\ell(V_2)=20$.\\
		
\noindent On the other hand, it is easy to check that the final matrix 
$M= \left( \begin{array}{cc} a_{t-1} & b_{t-1} \\ 
	a_t & b_t \end{array} \right)$ satisfies 
\noindent $M \left(\begin{array}{c} U\\ 
	V \end{array} \right) = 
\left( \begin{array}{cc} -62729 & 81769 \\ 
	353414 & -460685 \end{array} \right) \times
\left( \begin{array}{c} 922375420941 \\ 
	7075 9930  7587 \end{array} \right) =		
\left( \begin{array}{c} 1873414  \\ 
	725479 \end{array} \right) $.\\	
		
\noindent Now, if $u$ and $v$ were larger in size, namely:\\

$\left( \begin{array}{c} u \\ 
	v \end{array} \right) =	
\left( \begin{array}{c} 879645 ~\bullet ~785 421 
~\bullet u_3 ~\bullet u_4~ \ldots ~\bullet u_k\\ 
	674819 ~\bullet ~299 843 
	~\bullet v_3 ~\bullet v_4~ \ldots ~\bullet v_k\\ 
	\end{array} \right)$,\\

\noindent then we have 
to continue the half-gcd process. Since $W$ bits have been already 
chopped,  we have to update only the double, 
i.e.: multiply the next $2W$ leading bits of $U$ and $V$ by $M$, 
namely perform:\\

$ \left( \begin{array}{c} U_3 \bullet U_4 \\ 
	V_3 \bullet V_4 \end{array} \right) ~ \leftarrow M \times 
\left( \begin{array}{c} U_3 \bullet U_4 \\ 
	V_3 \bullet V_4 \end{array} \right)$, and disregard all the 
other bits of $U$ and $V$.\\

Then do the same process as before with 
$ \left( \begin{array}{c} U_2 \bullet U_3 \\ 
	V_2 \bullet V_3 \end{array} \right)$ instead of 
$ \left( \begin{array}{c} U_1 \bullet U_2 \\ 
	V_1 \bullet V_2 \end{array} \right)$ to chop the vector 
$ \left( \begin{array}{c} U_2  \\ 
	V_2  \end{array} \right)$, and so on, repeating the process 
	till we reach the middle of the size of $U$.\\


\section{An Example} \label{exa}

\noindent Let us consider two consecutive Fibonacci numbers 
$(u,v) = (F_{59},F_{58})$, i.e.:\\
    $
\left( \begin{array}{c} F_{59} \\ 
	F_{58} \end{array} \right) =
\left( \begin{array}{c} 956722026041 \\
	591286729879 \end{array} \right) = $  	
$		
\left( \begin{array}{c}  956722 \\ 
  591286 \end{array} \right)  10^{9}+
\left( \begin{array}{c} 026041     \\ 
	729879 \end{array} \right)= 		
\left( \begin{array}{c} 956722 ~\bullet ~026041 \\ 
 591286 ~\bullet ~729879 \end{array} \right)$.\\
First, we must compute $mMAX$ which gives $2^{mMAX}$, the maximum size of 
the output matrix $L_1$
$$mMAX = \ell(v) - \lceil \frac{\ell(u)}{2} \rceil -1 = 19.$$	
	
Let $(U_1,V1)= (956722,591286)   $ then $\ell(U_1)= \ell(V_1)=20$, 
$n=p=20$ and $m=p-1 - \lceil \frac{n}{2}  \rceil =9 $. 
We run the Extended Euclid algorithm and stops 
 at $|a_i| \leq 2^9$. We obtain the matrix $L_0$ and the remainders 
 ($r_1,r_2$):\\
\hspace*{1.5cm} $L_0=ILE(U_1,V_1)=   
	\left( \begin{array}{cc} -144 & 233 \\ 
	233 & -377 \end{array} \right)$ \  and \ 
	$(r_1,r_2)= (1670,1404)$. \\	
\hspace*{1.5cm} Hence $ L_0 \left(\begin{array}{c} U\\ 
	V \end{array}\right) = 
L_0 \left(\begin{array}{c} U_1~10^{9}+U_2\\ 
	V_1~10^{9}+V_2 \end{array}\right) = 
 L_0 \left(\begin{array}{c} U_1\\ 
	V_1 \end{array}\right)10^{9}+
L_0 \left(\begin{array}{c} U_2\\ 
	V_2 \end{array}\right)$. \\
	\noindent 
By {\tt update($1,0$)} we compute 
$L_0 \left(\begin{array}{c} U_2\\ 
	V_2 \end{array}\right)   $, so \\
$L_0  
\left( \begin{array}{c} U_1~ 10^{9}+U_2 \\ 
	V_1~ 10^{9}+V_2 \end{array} \right) =	
\left( \begin{array}{c} 1670 \\ 
	1404 \end{array} \right) 10^{9} +
\left( \begin{array}{c} 166 311 903 \\ 
	-269 096 830 \end{array} \right) 		
= \left( \begin{array}{c} 1836 \\ 
	1134 \end{array} \right) 10^{9} +
\left( \begin{array}{c} 311 903 \\ 
	903 170 \end{array} \right)$,\\	
\noindent where {\tt FIX } occurs in the last equality, hence 
the new values of $U_1$  and $V_1$:\\
\hspace*{2cm} $ \left( \begin{array}{c} U_1 \\ 
	V_1 \end{array} \right)	\longleftarrow 
\left( \begin{array}{c} 1836 \\ 
	1134 \end{array} \right)$ and 	
$ \left( \begin{array}{c} U_2 \\ 
	V_2 \end{array} \right)	\longleftarrow 
\left( \begin{array}{c} 311 903 \\ 
	903 170 \end{array} \right)$.\\

\noindent We apply {\tt ILE} to the new integers   
$U= 1 836 311$ and $V=1 134 903$. We have $\ell(U)= 21$ and $\ell(V)
=21$. Repeating the same process as before, we obtain $m=9$, 
the second matrix $R_0$ and remainders ($r_1,r_2$)\\
\hspace*{1.5cm} $R_0=ILE(U,V)=   
	\left( \begin{array}{cc} 233 & -377 \\ 
	-377 & 610 \end{array} \right) $ \ and \ 
	$(r_1,r_2) = (2032,1583)$.\\
Since $L_1 = R_0 \times L_0$ and using {\tt update}(2,0) we obtain \\		
\hspace*{1.5cm}  $ L_1 \left(\begin{array}{c} U\\ 
	V \end{array}\right) = 	
R_0 \left(\begin{array}{c} 1 836 311 \\ 
	1 134 903 \end{array}\right) 10^6 +  
 R_0 \left(\begin{array}{c} 903\\ 
	 170\end{array}\right)= 
\left(\begin{array}{c} 2032\\ 
	1583 \end{array}\right)10^6 +
\left(\begin{array}{c} 146 309\\ 
	-236 731\end{array}\right)$. \\	
\hspace*{1.5cm} Thus $L_1 \left(\begin{array}{c} U\\ 
	V \end{array}\right) = 	
\left(\begin{array}{c} 2 178 309\\ 
	1 346 269 \end{array}\right) \ = \ 
\left(\begin{array}{c} F_{32}\\ 
	F_{31} \end{array}\right)$. \\	
and $L_1= R_0 \times L_0= \left( \begin{array}{cc} 233 & -377 \\ 
	-377 & 610 \end{array} \right) \times 
\left( \begin{array}{cc} -144 & 233 \\ 
	233 & -377 \end{array} \right) =  			
\left( \begin{array}{cc} -121393 & 196418 \\ 
	196418 & -317811 \end{array} \right)$.\\
	
We can apply one more Euclid step because the matrix $Q \times L_1$ still 
satifies $|a_i| \leq 2^{mMAX} = 2^{19} = 524288$. So after one Euclid step 
we finally obtain:\\
\hspace*{3cm} $Q \times L_1 \left(\begin{array}{c} U\\ 
	V \end{array}\right) = 		
\left(\begin{array}{c} 1 346 269\\ 
	832 040 \end{array}\right) \ = \ 
\left(\begin{array}{c} F_{31}\\ 
	F_{30} \end{array}\right)$,  \\		
and after $L_1 = Q \times L_1$, we have:\\ 
\hspace*{4cm} $L_1 = 	 			
\left( \begin{array}{cc} 196418 & -317811 \\ 
-317811 & 514 229 \end{array} \right)$.\\

\noindent Moreover, at this level, we may consider that $U_1$ and $V_1$ 
are 
eliminated (chopped), even if $U_2$ has some extra bits,  and that $U_2$ 
and $V_2$ are updated as follows:\\

$ \left( \begin{array}{c} U_2 \\ 
	V_2 \end{array} \right)	= 
\left( \begin{array}{c} 1873414  \\ 
	725479 \end{array} \right) = 
\left( \begin{array}{c} 1 \bullet 824838  \\ 
	0 \bullet 725479 \end{array} \right) $;
	 \ \ $\ell(U_2)=21$; \ \ $\ell(V_2)=20$.\\
		
\noindent On the other hand, it is easy to check that the final matrix 
$M= \left( \begin{array}{cc} a_{t-1} & b_{t-1} \\ 
	a_t & b_t \end{array} \right)$ satisfies 
\noindent $M \left(\begin{array}{c} U\\ 
	V \end{array} \right) = 
\left( \begin{array}{cc} -62729 & 81769 \\ 
	353414 & -460685 \end{array} \right) \times
\left( \begin{array}{c} 922375420941 \\ 
	7075 9930  7587 \end{array} \right) =		
\left( \begin{array}{c} 1873414  \\ 
	725479 \end{array} \right) $.\\	
		
\noindent Now, if $u$ and $v$ were larger in size, namely:\\

$\left( \begin{array}{c} u \\ 
	v \end{array} \right) =	
\left( \begin{array}{c} 879645 ~\bullet ~785 421 
~\bullet u_3 ~\bullet u_4~ \ldots ~\bullet u_k\\ 
	674819 ~\bullet ~299 843 
	~\bullet v_3 ~\bullet v_4~ \ldots ~\bullet v_k\\ 
	\end{array} \right)$,\\

\noindent then we have 
to continue the half-gcd process. Since $W$ bits have been already 
chopped,  we have to update only the double, 
i.e.: multiply the next $2W$ leading bits of $U$ and $V$ by $M$, 
namely perform:\\

$ \left( \begin{array}{c} U_3 \bullet U_4 \\ 
	V_3 \bullet V_4 \end{array} \right) ~ \leftarrow M \times 
\left( \begin{array}{c} U_3 \bullet U_4 \\ 
	V_3 \bullet V_4 \end{array} \right)$, and disregard all the 
other bits of $U$ and $V$.\\

Then do the same process as before with 
$ \left( \begin{array}{c} U_2 \bullet U_3 \\ 
	V_2 \bullet V_3 \end{array} \right)$ instead of 
$ \left( \begin{array}{c} U_1 \bullet U_2 \\ 
	V_1 \bullet V_2 \end{array} \right)$ to chop the vector 
$ \left( \begin{array}{c} U_2  \\ 
	V_2  \end{array} \right)$, and so on, repeating the process 
	till we reach the middle of the size of $U$.\\


\noindent Here we stress that this is the main difference between our 
approach 
and the other Sorenson's like algorithms,  where 
{\bf all} the bits of $U$ and $V$ are updated by 
multiplying them with the matrix $M$. In our approach, 
we only update the double of 
what we have already chopped.\\ 

\section{Remarks} \label{conc}
Unlike the recursive versions of 
GCD algorithms (($[1]$),$[4]$), our aim is not to balance 
 the computations on each leaf of the binary tree
computations, but to make full of single precision everytime, computing therefore
the maximum of quotients in single precision. This lead to different
computations each step in the algorithm AEA and the other recursive GCD algorithms. 
Moreover the fundamental difference between AEA and Sch\"onhage's approach 
is that AEA deals straightforward with the most significant leading bits 
first ({\bf MSF computation}).
Consequently, 
we can stop the algorithm AEA at any time and still obtain the leading 
bits of the result. Thus AEA is a strong MSF algorithm and can be 
considered for ''on line" arithmetics, where all the basic operations 
can be carried out simultaneously as soon as only the needed bits are available.  
This new algorithm should be an alternative to the Sch\"onhage GCD algorithm.  
On the other hand, the derived GCD algorithm deals with many applications where 
long euclidean divisions are needed. 
We have identified and started to study 
many applications such as subresultants and Cauchy index computation, 
$Pade$-approximates or $LLL$-algorithms.


\section{References}

 
$[1]$ ~{\sc Gathen, J. von zur, Gerhard, G.}
\newblock Modern Computer Algebra.
\newblock In {\em Cambridge University Press}
  (1999). \\
$[2]$ ~{\sc D.H.~Lehmer.} 
\newblock{Euclid's algorithm for large 
numbers,}
\newblock{American Math. Monthly,}
\newblock{45, 1938, 227-233.} \\
$[3]$ ~{\sc M.F.~Roy and S.M.~Sedjelmaci.} 
\newblock{The Polynomial 
Accelerated Euclidean Algorithm, Subresultants and Cauchy index,}
\newblock{work in progress,~}{2004.} \\
$[4]$ ~{\sc A.~Sch\"onhage.} 
\newblock{Schnelle Berechnung von 
Kettenbruchentwicklugen,}
\newblock{Acta Informatica,}
\newblock{1, 1971, 139-144.} \\
$[5]$ ~{\sc A.~Sch\"onhage, A.,F.,W. Grotefeld and E. Vetter.}
\newblock{Fast Algorithms, 
A Multitape Turing Machine Implementation,}
\newblock{BI-Wissenschaftsverlag,}
\newblock{Mannheim, Leipzig, Z\"urich, 1994.} \\
$[6]$ ~{\sc A.~Sch\"onhage, V. Strassen.}
\newblock{Schnelle Multiplication grosser Zalen}
\newblock{Computing 7,}{1971, 281-292.} \\
$[7]$ ~{\sc S.M.,~Sedjelmaci.} 
\newblock{On A Parallel Lehmer-Euclid 
GCD Algorithm,}
\newblock{in Proc. of the International Symposium on 
Symbolic and Algebraic 
Computation (ISSAC'2001),}
\newblock{2001, 303-308.}\\

\vspace*{1cm}
\noindent Sidi Mohamed Sedjelmaci\\
LIPN CNRS 
UMR 7030, \\
Universit\'e Paris-Nord\\
Av. J.-B. Cl\'ement, 93430 Villetaneuse, 
France. \\
E-mail: 
$sms$$\char64$lipn.univ-paris13.fr

\end{document}